# Uniform vapor pressure based CVD growth of MoS$_2$ using MoO$_3$ thin film as a precursor for co-evaporation

Sajeevi S. Withanage,[†,‡] Hirokjyoti Kalita,[‡,∥] Hee-Suk Chung,[⊥] Tania Roy,[‡,§,∥] Yeonwoong Jung,[‡,§,∥] Saiful I. Khondaker*[,†,‡,∥]

[†] University of Central Florida, Department of Physics, 4111 Libra Drive, Physical Sciences Bldg. 430, Orlando, FL 32816, United States

[‡] University of Central Florida, NanoScience Technology Center, 12424 Research Pkwy #400, Orlando, FL, United States

[§] University of Central Florida, Department of Materials Science & Engineering, 12760 Pegasus Drive, Engineering I, Suite 207, Orlando, FL, 32816, United States

[∥] University of Central Florida, Department of Electrical & Computer Engineering, 4328 Scorpius Street, Orlando, FL 32816, United States

[⊥] Korea Basic Science Institute, Analytical Research Division, Geonji-road 20, Jeonju, 54907, South Korea

**Abstract:**

Chemical vapor deposition (CVD) is a powerful method employed for high quality monolayer crystal growth of 2D transition metal dichalcogenides with much effort invested toward improving the growth process. Here, we report a novel method for CVD based growth of monolayer molybdenum disulfide (MoS$_2$) by using thermally evaporated thin films of molybdenum trioxide (MoO$_3$) as the molybdenum (Mo) source for co-evaporation. Uniform evaporation rate of the MoO$_3$ thin films provides uniform Mo vapor which promotes highly reproducible single crystal growth of MoS$_2$ throughout the substrate. These high-quality crystals are as large as 95 μm and were characterized by scanning electron microscopy, Raman spectroscopy, photoluminescence spectroscopy, atomic force microscopy and transmission electron microscopy. The bottom gated field effect transistors fabricated using the as grown single crystals show n-type transistor behavior with a good on/off ratio of $10^6$ under ambient condition. Our results presented here addresses the precursor vapor control during the CVD process and is a major step forward toward reproducible growth of $MoS_2$ for future semiconductor device applications.

**Introduction:**

Layered two-dimensional (2D) semiconducting transition metal dichalcogenides (TMDCs) ($MX_2$: M=Mo, W, X=S, Se, Te) have attracted a great deal of attention due to their exciting properties emerged from quantum confinement effects including intrinsic direct bandgap in single layer, high photoconductivity[1-2], appreciable charge mobility and high on/off ratio[3]. In particular, 2D molybdenum disulfide ($MoS_2$) has been studied extensively over the past few years because of its natural abundance and making expedited progress in its applications in the fields of electronics and optoelectronics with photodectors[4], light emitting devices[5], solar cells[6] and bio sensors[7-9]. Most of these studies were done in exfoliated samples, which limits their practical application due to the low yield, small lateral size and uncontrolled layer number. For scalable device fabrication, large scale synthesis of $MoS_2$ is highly desirable. Chemical vapor deposition (CVD) based synthesis methods including thermal vapor sulfurization of $Mo/MoO_3$ films[10-12], thermal decomposition of thiosalts[13] and gas phase reaction of $MoCl_5$ and $H_2S$ gas[14] has been employed for large scale synthesis however, obtaining $MoS_2$ samples with controlled layer number and high electrical and optical quality remains elusive with these methods. The co-evaporation of Mo and S precursors based CVD has received much attention for the past few years due to its ability to produce high quality MoS$_2$ single crystals with excellent electrical and optical properties[15-18]. The challenging nature of CVD growth of MoS$_2$ continues to stimulate intense research effort in this field. If the CVD is going to be used for large scale production of MoS2 samples, it is important to understand how small variations of CVD process parameters could lead to different results[19-20].

In the co-evaporation based CVD technique MoO$_3$ powder and S powder are commonly used to produce Mo and S vapor which then reacts to produce MoS$_2$ crystal following the two steps chemical reactions[21].

$$MoO_3 + \frac{x}{2}S \rightarrow MoO_{3-x} + \frac{x}{2}SO_2$$

$$MoO_{3-x} + \frac{7-x}{2}S \rightarrow MoS_2 + \frac{3-x}{2}SO_2$$

Since most of the growth-related reactions occur in the vapor phase, the uniform melting/sublimation of the precursors and maintaining uniform vapor pressure of the gas phase precursors is highly desirable for clean and reproducible growth of MoS$_2$. However, such a control of uniform vapor pressure turned out to be very challenging resulting in a growth of oxide/oxysulfide (MoO$_2$/MoOS$_2$) species along with MoS$_2$[19, 22]. To our knowledge, the importance of controlling sulfur vapor pressure is discussed in detail in the literature[23-24], but discussion of controlling Mo vapor pressure is rare. In particular, having a uniform vapor pressure of Mo in the reaction region when sulfur enters can help to maintain proper Mo:S ratio for the successful and reproducible growth of MoS$_2$. However, when small amount of MoO$_3$ powder is used, due to the different particle sizes in the powder, MoO$_3$ cannot be uniformly distributed in the crucible which might hinder clean and reproducible growth of MoS$_2$.

Here we show that by replacing MoO$_3$ powder with a MoO$_3$ thin film for Mo precursors, clean and reproducible growth of monolayer MoS$_2$ can be obtained. MoO$_3$ thin film of 5 – 20 nm was thermally evaporated from MoO$_3$ powder which was then used as the Mo precursor for co-evaporation. Uniform evaporation rate of our thin film precursor facilitates better control of Mo vapor in the vapor phase resulting in clean triangular crystals of monolayer MoS$_2$ on Si/SiO$_2$ substrates with high crystal quality and uniformity proved by scanning electron microscopy (SEM), Raman spectroscopy, photoluminescence (PL) spectroscopy, atomic force microscopy (AFM) and transmission electron microscopy (TEM). Electric characterization of these pristine crystals with bottom gated field effect transistors (FETs) showed n type transistor behavior with a good on/off ratio of 10$^6$ under ambient conditions verifying the ability of these materials in modern semiconductor electronics

**Results:**

We synthesized MoS$_2$ single crystals in a home-built atmospheric pressure CVD system. Figure 1a) shows our experimental setup for this study. MoO$_3$ thin films were deposited on Si/SiO$_2$ substrate using thermal evaporation and used as the Mo source for co-evaporation synthesis of MoS$_2$ crystals in a single zone tube furnace. Substrates were placed faced down toward the thin film source. Sulfur was placed upstream at a low temperature zone. The temperature profiles for MoO$_3$ and S are shown in figure 1c). Further information on the growth process can be found in the methods section.

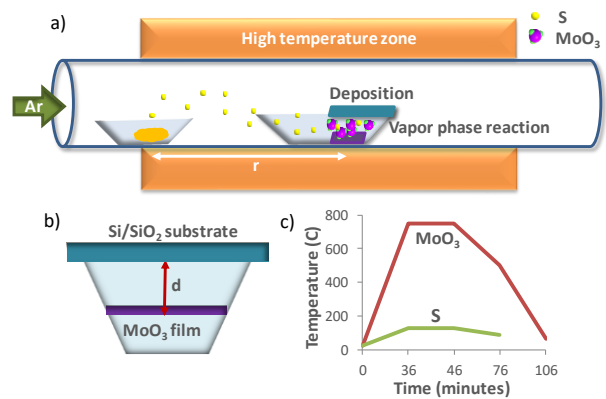

**Figure 1**. Experimental setup. a) schematic representation of the atmospheric pressure CVD setup and the relative S, MoO$_3$ and substrate positioning. b) Cross section view of the substrate boat: target substrate was placed face down toward the film at a small distance d. c) The temperature profile of MoO$_3$ and S at setpoint value of 750 $^0$C.

Figure 2 shows a representative result for our optimized CVD growth of MoS$_2$. For this, we used 20 nm thin film of MoO$_3$ as a precursor for Mo vapor. S was placed at the edge of the furnace, 23.5 cm upstream from the center of the furnace and the growth temperature was 750 °C. Figure 2(a) show an optical image of the whole Si/SiO$_2$ substrate after MoS$_2$ growth as well as a bare substrate for comparison. The darker region of the growth substrate is where the crystals are grown. This region of the substrate is directly facing toward the thin film MoO$_3$ source. To show the growth on this substrate in detail, we divided this region into nine sections as shown in figure 2b) and present optical micrographs of each of these region (labelled 1 - 9) in figure 2(d). Form here, we see that dense crystal grains of self-seeded $MoS_2$ are grown throughout the growth substrate without any traces of $MoO_2/MoOS_2$. This is in clear contrast to what we have observed when we used a small amount of (~1 mg) $MoO_3$ powder as a Mo precursor where we saw isolated regions of $MoS_2$ growth along with rhomboidal $MoO_2/MoOS_2$ on $MoS_2$ in other regions (figure S1). Increasing amount of $MoO_3$ powder only increased the regions of $MoO_2/MoOS_2$. Since the difference between the two scenarios is the uniform sublimation of $MoO_3$ thin film which causes uniform evaporation of Mo in the growth region, we conclude that uniform evaporation of Mo has an important role in dense singe crystal growth of $MoS_2$. We repeated the $MoO_3$ thin film growth over 50 times and we always obtained clean $MoS_2$ growth without any oxysulfides. This was not the case when we used $MoO_3$ powder, where clean and reproducible $MoS_2$ crystals were hard to achieve. Another important observation is that all the single crystals are triangular in shape. Given the conditions of the growth and the rough edges of these triangular crystals (Figure 3b) which are S-zz terminations, we conclude that they are grown at a sulfur sufficient environment[24-25]. The shape and size variation of $MoS_2$ crystals has been discussed in the literature using both experiment and theory[25-27]. Depending upon the Mo vapor concentration with respect to the S, the $MoS_2$ crystals can show different shapes (hexagon, truncated triangle, and triangle), sizes, and density variation as well as formation of $MoO_2/MoOS_2$ structures. Wang. et al.[26] observed variations in the $MoS_2$ crystal shape from triangular to hexagonal geometries with respect to the distance from $MoO_3$ powdered precursor location. This variation in crystal shape was attributed to the variation of Mo vapor concentration along the growth substrate. Wu et al[27] also observed shape and size variation of $MoS_2$ with respect to $MoO_3$ concentration. These studies suggest that, there is a small window of Mo:S vapor ratio (both in Mo reach and S rich region) during which $MoS_2$ crystal can maintain triangular shape although their size could vary slightly due to a slight variation of Mo:S vapor ratio. With a relatively larger variation of vapor ratio, both shape and size can vary. While too much Mo vapor variation could lead to $MoS_2$ in some places while $MoO_2/MoOS_2$ in other places. When $MoO_3$ powder was used as a precursor in our experiment (figure S1), there appears to be a large variation of Mo:S vapor pressure throughout the growth region. On the other hand, when $MoO_3$ thin film was used as precursor for Mo vapor, we observe same triangular crystal shape throughout the growth region (no shape variation, neither any oxysulfide species) suggesting that we are able to maintain a tight control on Mo vapor concentration in the gas phase by replacing the powdered precursor by the thin film precursor.

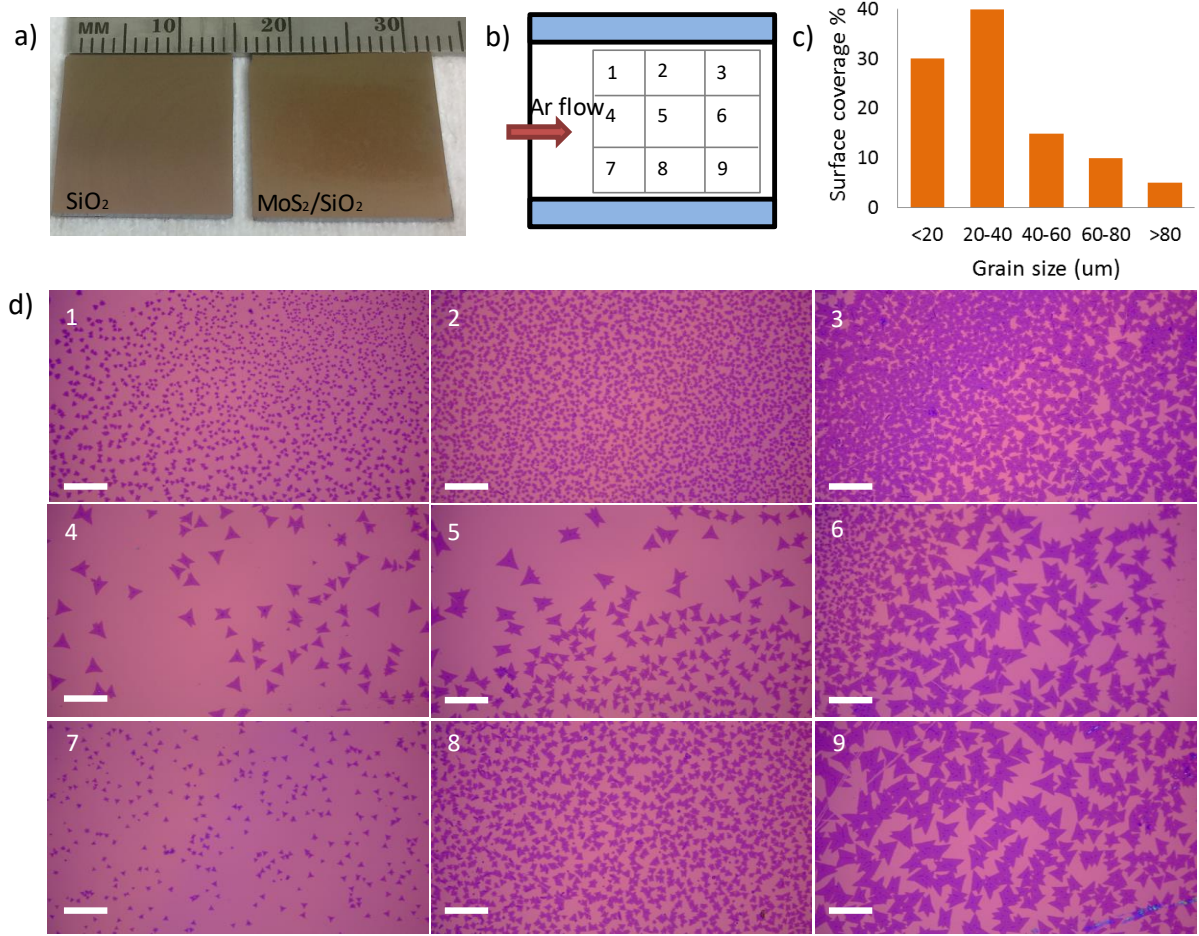

**Figure 2.** Growth Results. a) Contrast between bare Si/SiO$_2$ substrate and substrate after the growth of MoS$_2$. b) A sketch of the sample with the corresponding image positions. The blue regions marked here are outside of the boat. c) The grain size histogram with the surface coverage. d) the optical images of the source at the positions specified in b). Scale bar is 100 µm in each image.

Figure 2(d) also show a size and density variation of the $MoS_2$ grains in different regions. A statistical view to the grain size variation is provided in figure 2c). We observed crystal sizes up to 95 µm which is the largest we measured for this growth and most crystals favored to grow in 20-40 µm size that covered around 40% of the growth region. The size variation and increased nucleation density is consistent with the picture of maintaining a tight control on Mo vapor concentration discussed above. A slight spatial variation of Mo:S vapor ratio due to a small variation of either Mo or S or both dictated by the flow, temperature gradients and diffusion patterns within the gas can give rise to different sizes of $MoS_2$ crystals while maintaining triangular shape even through Mo evaporation is homogenous. In some regions, the triangular domains are merged to form polycrystalline aggregates[28] can be seen due to the random orientations of the crystals and higher density[28-31]. Collectively the absence of any oxysulfide phase along with the ability of maintaining triangular shape validates our capability to maintain a controlled Mo vapor condition for the clean growth of $MoS_2$.

In addition to MoS$_2$ grown on the growth substrate, we have occasionally observed scattered MoS$_2$ growth on source substrate as well (Figure S7), although the coverage density is much lower in comparison to the faced down growth substrate. This is an interesting observation given the unique geometry of our growth setup where the Mo and S vapors are trapped between two parallel plates the gas flow patterns also differ to the conventional face down technique where the growth substrate sits on top of a crucible with MoO$_3$ powder precursor. We note that, a continuous coverage of monolayer MoS$_2$ was not obtained using our growth method neither on the growth substrate, nor on the source substrate. The MoS$_2$ single crystal growth reported here using MoO$_3$ thin film as an evaporation source is different than the sulfurization of Mo thin films previously reported by us and other groups[32-33]. In the present work, the reaction between Mo and S occurs in the vapor phase unlike the sulfurization of Mo thin films where layer by layer sulfurization occurs at the thin film surface. In the present work, S was introduced to the reaction zone when the center of the furnace was at 650 °C. This is to ensure that all MoO$_3$ thin sublimated before S enters the reaction zone. In a control experiment (no S, all other parameters are identical to the optimized growth process), we found that 10 nm thin MoO$_3$ thin film completely gone from the source substrate at a temperature of 550 °C (in atmospheric pressure). The resulting MoS$_2$ crystal qualities are also different as the grain size in the present work is tens of microns while in the case of thin film sulfurization, the grain size is in tens of nanometer. As a result, the MoS$_2$ crystals grown using this method shows very high crystal quality, uniformity and better device properties discussed below compared to the polycrystalline films obtained by thin film sulfurization methods.

Figure 3a) shows an optical micrograph of a single crystal grown with the size ~95 μm. These big crystals are mostly composed of monolayer regions along with some isolated dendritic bilayers[34-36] region which can be clearly seen in the SEM image as well (figure 3b, spot 2). We only observe this bi-layer formation in large size crystals which could be due to the added nucleation sites by the defects in the first layer[36]. Self-seeded nucleation center is also visible in the SEM image (figure 3b, spot 1). The tapping mode AFM image (figure 3c) taken at a single crystal edge was analyzed to identify the number of layers of the grown crystals. The step height at this edge was measured to be 0.78 nm which corresponds to the monolayer $MoS_2$ thickness and the topography of the crystal shows a highly uniform layer. Raman single spectra (figure 3d) was acquired to confirm the chemical composition of the material grown as well as the layer number. As shown in figure 3c, Raman spectrum has the two prominent peaks, $E^1_{2g}$ and $A_{1g}$ which corresponds to in plane and out of plane vibrations of Mo and S atoms with respective wave numbers of 390.76 cm$^{-1}$ and 410.76 cm$^{-1}$. Spacing between these two peaks (Δ) is 19.82 cm$^{-1}$ conistent with monolayer $MoS_2$[37-38].

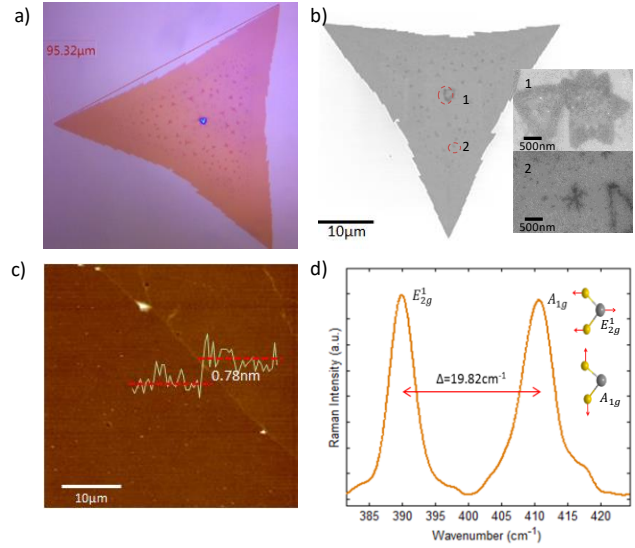

**Figure 3.** Characterization of the $MoS_2$ single crystals. a) optical image of one of the largest monolayer crystals observed. b) SEM image of a single crystal. c) The AFM topography and height profile (inset) taken at an edge of a monolayer domain. d) $E^1_{2g}$ and $A_{1g}$ vibrational modes of atoms and Raman single spectra for a single crystal.

Figure 4(a) shows a PL spectrum from a representative single crystal with two signature peaks at 1.85eV and 1.99eV which corresponds to A (direct band gap excitations) and B (excitations resulted from valence band splitting due to strong spin-orbit coupling[39]) direct excitronic transitions[37, 40-41]. The full width at half-maximum (FWHM) of the A peak is measured to be

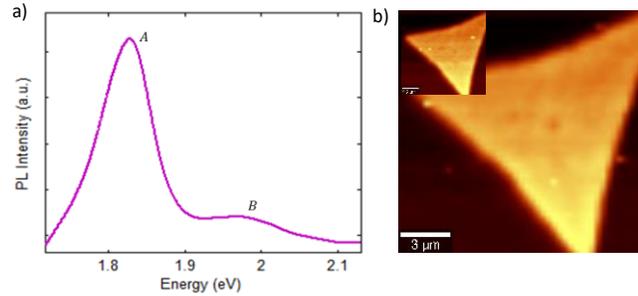

**Figure 4.** PL characterization. a) PL single spectra of a $MoS_2$ crystal grown with the optimized growth recipe. b PL intensity mapping of A peak for the same crystal. Inset shows mapping of B peak

~80 meV which is narrower than the exfoliated monolayer $MoS_2$ samples on $Si/SiO_2$ substrates[39, 42] showing better optical quality of our samples. The even color contrast of PL intensity mapping of A and B peaks (figure 4b) confirms the uniform layer thickness of the crystal.

Scanning transmission electron microscopey (STEM) was employed to further varify the crystal quality of the $MoS_2$ crystals. As grown $MoS_2$ crystals were transferred to a TEM grid with a carbon mesh using buffered oxide etchant (details in methods). Figure 5 shows dark-field STEM micrographs for the monolayer $MoS_2$ grown using the thin film based co-evaporation. High resolution STEM (HRSTEM) image (figure 5b) clearly shows the high crystalline quality of the of the monolayers with periodic hexagonal atomic arrangement. The lattice constant is measuresd as 0.27 nm well agrees with previously reported results[29, 43]. The monolayer can also be identified from the folding edge of the transferred film (figure 5c). Evidantly in the SEM image, these big

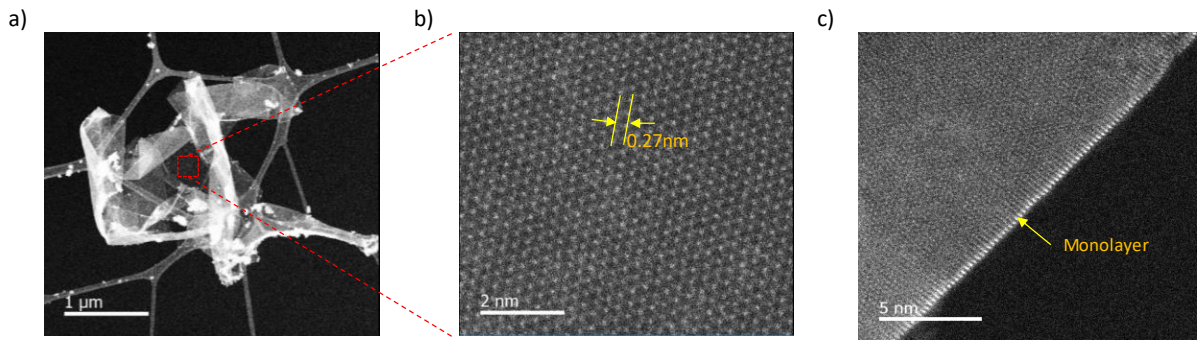

**Figure 5.** TEM characterization. a) low magnification DF-STEM image of the transferred thin film on the copper grid. b) HRSTEM image of the monolayer $MoS_2$ showing hexagonal atom arrangement. c) DF-STEM image for the folded edge of the monolayer

crystals shows some multilayer growth at some regions (nucleation center and ad-layers) and accordingly, the STEM data shows Moire patterns which caused by the diffraction due to the mis-allignment of these layers (Figure S5).

We also experimented this growth at higher temperatures. Figure S3 shows the optical micrographs of the crystals grown at 750 °C, 800 °C and 850 °C by using 20 nm $MoO_3$ thin films. We observe triangular shaped crystals at all growth temperature suggesting that uniform Mo vapor pressure can be maintained at elevated temperature. We also found that the coverage density generally decreases at higher growth temperatures. This could be linked to the reduced nucleation at higher temperatures. The sticking coefficient which defines the nucleation ability of the surface significantly reduces due to the enhanced desorption rate at the higher temperatures[44]. This phenomenon also explains the multilayer growth at the nucleation center at higher temperatures. Since there are less nucleation sites to facilitate the gas phase $MoS_2$ to adsorbed on to the substrate surface which results in growth of multiple $MoS_2$ layers at the same nucleation site. The multilayer growth at high temperatures is discussed as an effect of high precursor supply in the previous reports[31, 45-46] when $MoO_3$ powdered precursor is used which is not applicable for our thin film-based growth. In our experiments even at 750 °C, $MoO_3$ thin film precursor is completely evaporated and reacted to grow $MoS_2$, therefore more $MoO_3$ supply is not possible. The AFM,

Raman and PL characterizations for the crystals grown at higher temperature are shown in figure S4. The AFM topography shows a clear stepwise height variation and each step height measured is similar to monolayer MoS$_2$ thickness. Raman single spectra taken near the center of nucleation shows the $E^1_{2g}$ and $A_{1g}$ signature peaks for MoS$_2$ with the spacing of $\Delta = 23.43$ cm$^{-1}$ which corresponds to a few layer MoS$_2$[37, 46]. PL mapping of the crystals grown at 850 °C shows a clear PL intensity drop at these multilayers and very thick layers at the center shows no PL response. The effect of MoO$_3$ precursor amount is tested by varying the thickness of the film from 5nm to 20nm with the same amount of S (Figure S5). All the crystals grown have the similar shape suggesting that the uniform Mo pressure control is possible even with 5 nm thin film. However, the crystal sizes are reduced at lower thickness of MoO$_3$ due to the low precursor supply.

We have also measured electrical transport properties of MoS$_2$ samples in a botton gated field effect transistor (FET) geometry. FET devices were fabricated on as grown MoS$_2$ single crystals by using standard electron beam lithography (see methods). The nucleation center was excluded at the device fabrication to minimize the reflection of multiple layers in the transport measurements. We used nickel for source and drain electrodes. Linear I$_D$-V$_D$ confirm ohmic contacts are formed at the two electrodes and all the measurements were performed under ambient for as grown samples. Figure 6 shows electrical characteristics of a sample device (optical image is shown in the inset of figure 6b) . The field-effect mobility was extracted based on the slope $dI_D/dV_G$ fitted to the linear region of the $I_D$-$V_G$ using the equation $\mu = (L/WC_GV_D)(dI_D/dV_G)$, where L, W and $C_G$ are the channel length, channel width and and the gate capacitance. For the sample device shown in the figure, 5µm channel length and 6µm and channel width is used in mobility calculations. This device shows an n-type conduction with on/off ratio of $10^6$ and peak mobility ~4.5cm$^2$/Vs which is comparable with the mobilities for pristine devices fabricated on CVD grown monolayers in previous reports[28].

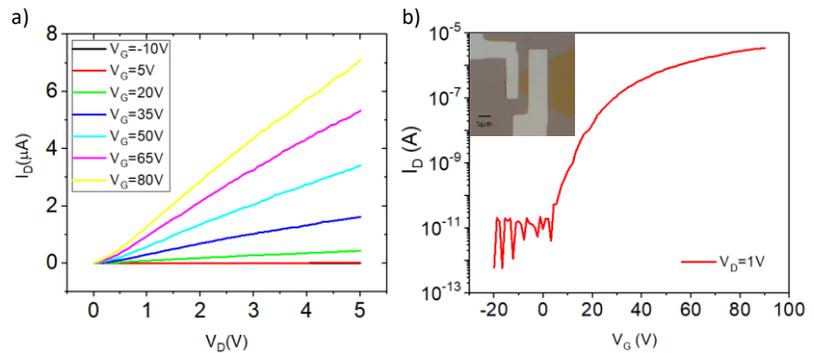

**Figure 6**. Electrical characteristics of as-grown MoS$_2$ single crystals. a) Output characteristics of the transistor device by sweeping the gate voltage (V$_G$) from -10V to 80V. b) Transfer characteristics. Plots are provided for the device shown in inset of b)

**Conclusions:**

In conclusion, we introduced a new CVD based method to synthesize monolayer $MoS_2$ crystals by using $MoO_3$ thin films as a precursor for co-evaporation. This new method facilitates uniform vapor pressure of Mo resulting in clean single crystal growth of $MoS_2$ throughout the substrate and are highly reproducible. As grown $MoS_2$ crystals show excellent Raman and PL response corresponding to high quality single crystal. TEM characterization further confirms crystalline quality of the sample. Room temperature electrical transport measurements show a mobility of ~5 cm$^2$/Vs with a current on/off ratio of $10^6$. We believe this new CVD process can be used for the synthesis of other TMDC materials and the growth can be extended to wide range of substrate types.

**Materials and Methods:**

*Growth process:* Si substrates with 250 nm thick oxide layer were used as the growth substrates as well as the substrates to deposit the $MoO_3$ thin films which was used as precursor for co-evaporation. The substrate was cleaned via sonication in acetone for 5 min followed by sonication in IPA for 5min, DI water rinse and 10 min mild oxygen plasma treatment. $MoO_3$ thin films with thickness 5-20 nm were deposited using thermal evaporation of $MoO_3$ powder (99.5%, Sigma

Aldrich) in a vacuum chamber at low evaporation rates of 0.2-0.5 A/s. The thickness of these thin films was confirmed via AFM topography (Figure S6). The thin film deposited substrates were cut into 1cm x 1cm size and placed in a ceramic boat, bare target substrates were placed on the same boat facing down with the separation (figure 1b) of 3-4 mm toward the source and placed at the center of the furnace. While the front end of the boat left open for sulfur vapor to enter the reaction zone easily, the rear end was covered by the substrate to trap the vapor phase precursors near the substrate, but a small spacing (less than a mm) left at the end to ensure sulfur vapor pressure doesn't overstep the balance between precursors. Ceramic crucible containing 600 mg of S powder (99.5%, Sigma Aldrich) was loaded in to a 1-inch quartz tube placed at the edge of the tube furnace (Barnstead International F79300 Tube Furnace) upstream. For different temperature setpoints, the $MoO_3$ and S precursor distance (r) is adjusted accordingly in such a way that S starts melting below 100 degrees from the setpoint value. Argon gas (99.995% purity) was used as the carrier gas. Temperature of the furnace was raised to the growth temperature of 750 - 850 °C at the rate of 20C/min and hold for 10 min (dwell time). After the dwell time the furnace could cool down naturally until the temperature dropped to 500 °C at which point, the furnace hood was opened for rapid cooling. 200sccm Ar flow was passed initially before heating up the furnace for 10 min to saturate the environment with argon, after which the flow rate was kept constant at 10 sccm. For the parameters used in this process, we found that 10 sccm gas flow with 10 min dwell time is optimum for our growth. When temperature of the furnace dropped to 350 °C, it was purged with 200sccm Ar again to flush off the excess reactants.

*Characterization of the materials:* The surface imaging and the grain size measurements were carried out by Olympus BX51M microscope equipped with Jenoptic Progres Gryphax camera) and The SEM images were taken with a Zeiss ULTRA-55 FEG SEM. A tapping mode AFM (Veeco instruments, Dimension 3100) topography was used to determine the height of $MoS_2$ crystals. The samples were scanned at slow rates in small scan areas at the edge of $MoS_2$ crystal to achieve more accurate height profile. Raman and PL spectroscopy measurements were carried out with confocal Raman microscope (Witec alpha 300 RA) at an excitation wavelength of 532 nm and with laser power of 0.293 mW at ambient conditions. For TEM characterizations; a droplet of buffered oxide etchant (BOE) was placed on the $Si/SiO_2$ substrate where $MoS_2$ crystals were grown. After a few minutes the oxide layer was etched, and the crystals were transferred to the liquid. The liquid was then picked up by a clean syringe and released on a TEM grid with a carbon mesh. All TEM/STEM measurements were performed with a JEOL ARM200F FEG-TEM/STEM with a Cs-corrector.

*Device fabrication and transport measurements:* The bottom gated FETs for the transport measurements were fabricated using standard electron beam lithography (EBL) using a Nabity Pattern Generator System connected to a Zeiss ultra 55 SEM for selected single crystals. Poly (methyl methacrylate) (PMMA) was spin coated on to the substrate containing $MoS_2$ samples and baked at 130 °C for 3 mins. The source and drain electrodes were then defined using EBL and developed in a solution of methyl isobutyl ketone and isopropyl alcohol. A 30 nm thick Ni film were deposited for the metal contacts using E-beam evaporation. The electron transport measurements of the final $MoS_2$ devices were carried out in a room temperature probe station using a Keysight Semiconductor Device Analyzer.

**Supporting Information:**

Additional information on growth results for co-evaporation of $MoO_3$ and sulfur powders, lateral substrate placement growth, growth at different temperatures, characterization of multilayer growth at higher temperature, effect of $MoO_3$ film thickness, supplementary TEM images, $MoS_2$ growth on source substrate, and AFM characterization of thin film MoO$_3$ source are provided in the supporting information

**Author Information:**

E-mail: Sajeevi S. Withanage <sajeevi@knights.ucf.edu>

E-mail: Hirokjyoti Kalita <hkalita@knights.ucf.edu>

E-mail: Hee-Suk Chung <hschung13@kbsi.re.kr>

E-mail: Tania Roy <tania.roy@ucf.edu>

E-mail: Yeonwoong Jung <yeonwoong.jung@ucf.edu>

E-mail: Saiful I. Khondaker <saiful@ucf.edu>

**Acknowledgement:**

This work was supported by U.S. National Science Foundation (NSF) under grants No. 1728309. We also acknowledge Dr. Tetard for Raman/PL spectroscopy support.

# Supporting Information

# Uniform vapor pressure based CVD growth of MoS$_2$ using MoO$_3$ thin film as a precursor for co-evaporation

*Sajeevi S. Withanage,[†,‡] Hirokjyoti Kalita,[‡,∥] Hee-Suk Chung,[⊥] Tania Roy,[‡,§,∥] Yeonwoong Jung,[‡,§,∥] Saiful I. Khondaker\*[†,‡,∥]*

[†] University of Central Florida, Department of Physics, 4111 Libra Drive, Physical Sciences Bldg. 430, Orlando, FL 32816, United States

[‡] University of Central Florida, NanoScience Technology Center, 12424 Research Pkwy #400, Orlando, FL, United States

[§] University of Central Florida, Department of Materials Science & Engineering, 12760 Pegasus Drive, Engineering I, Suite 207, Orlando, FL, 32816, United States

[∥] University of Central Florida, Department of Electrical & Computer Engineering, 4328 Scorpius Street, Orlando, FL 32816, United States

[⊥] Korea Basic Science Institute, Analytical Research Division, Geonji-road 20, Jeonju, 54907, South Korea

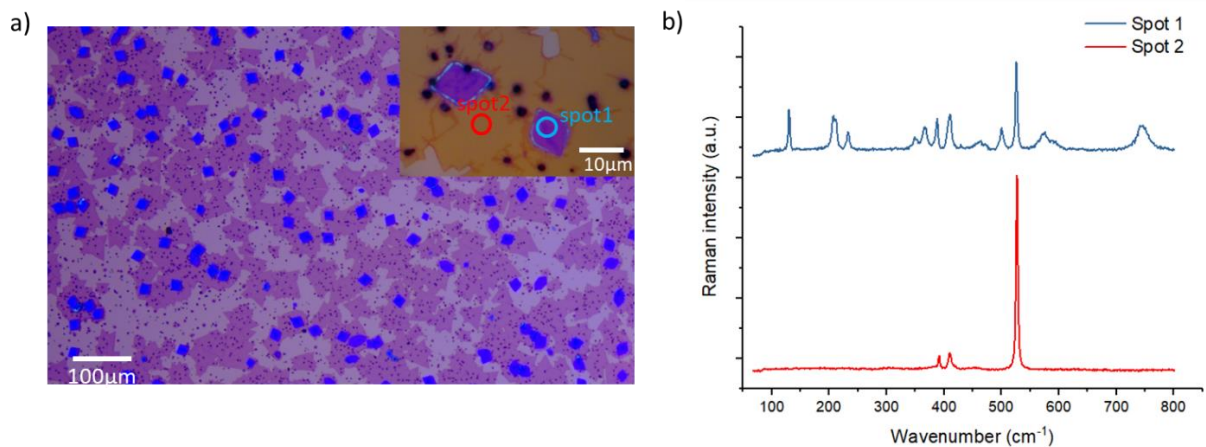

**Figure S1**. Growth results for co-evaporation of $MoO_3$ and sulfur powders. a) Optical micrographs for the crystals grown with ~1mg of $MoO_3$ powder and 600mg of S with the same setup. b) The Raman single spectra taken at spot 1 (blue) and spot 2 (red) shown in the inset of a). The Raman spectra is a combination at spot 1 is $MoS_2/MoOS_2$[1] and the reduced intensity of Si peak is an indication that these plates are very thick. Also, the high intensity of $MoS_2$ peaks implies that a thick layer of $MoS_2$ is grown on top of the oxysulfide plates. The spot 2 shows a $\Delta=18.92$ cm$^{-1}$ between $E_{2g}^1$ and $A_{1g}$ peaks which corresponds to monolayer $MoS_2$.

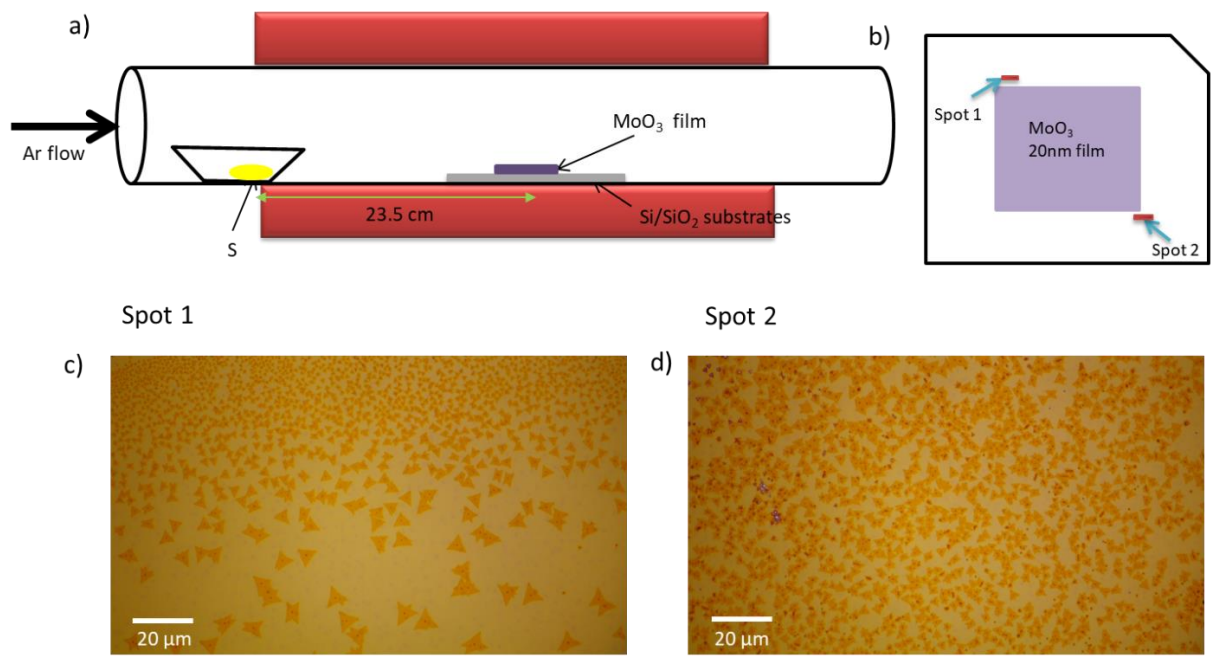

**Figure S2**. Lateral placement of source and substrate. a) Experimental setup showing relative positioning of substrate and precursors. b) Locations of MoS$_2$ growth on the substrate. c), d) Optical micrographs at spots 1 and 2 respectively. Using the same MoO$_3$ film thickness (20 nm) and same amount of S (600 mg), we observed that MoS$_2$ growth is limited to few spots near the source edges in lateral placement geometry.

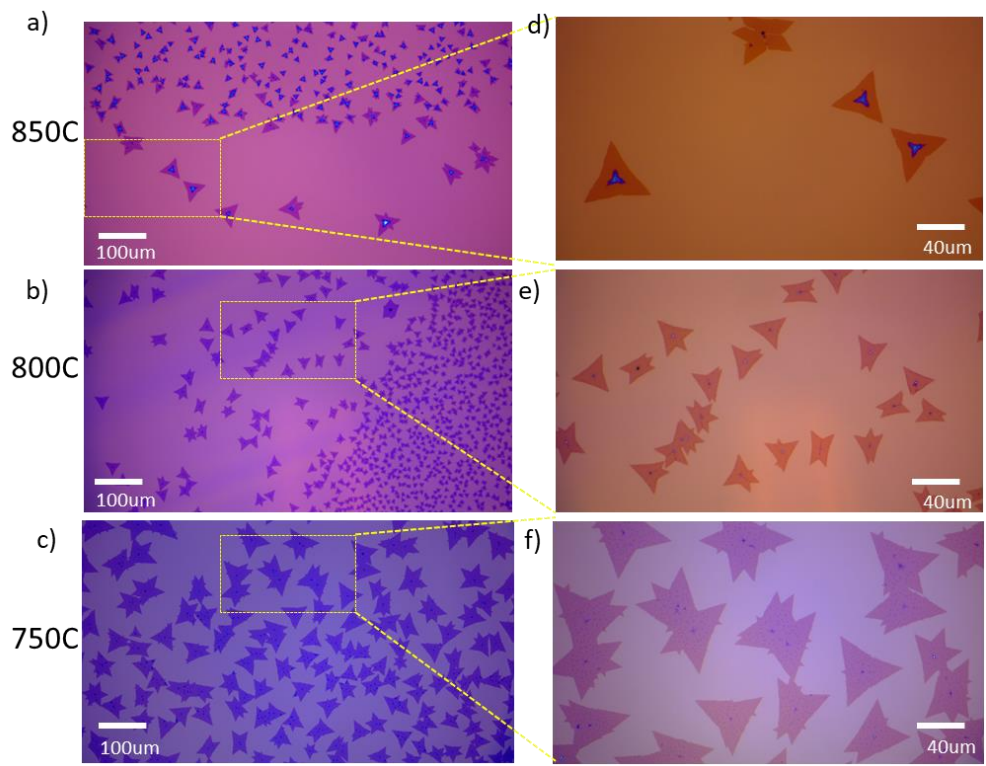

**Figure S3**. Growth at different temperatures. a)-c) Optical images of growth at 850C, 800C and 750C respectively taken at 20X magnification. d)- f) corresponding images at 50x magnification.

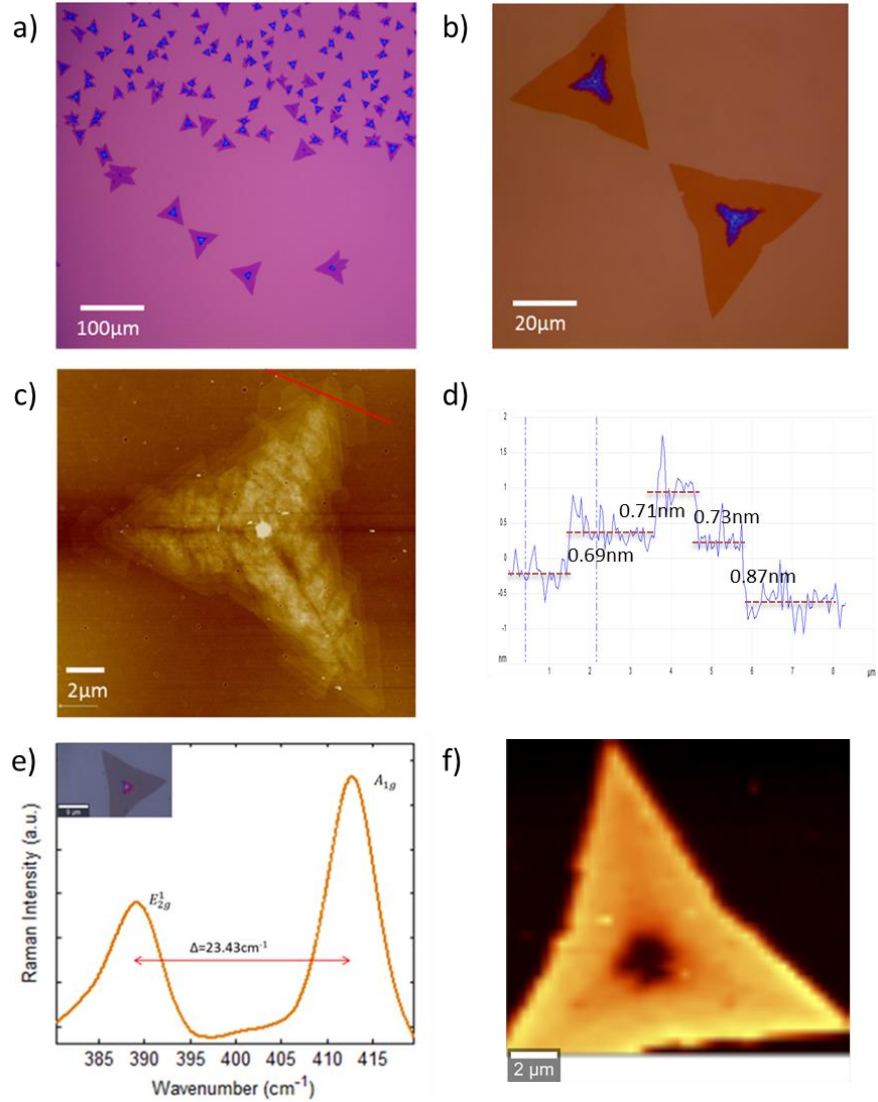

**Figure S4.** Multilayer growth around the nucleation center at 850°C. a), b) Optical images taken at 20x and 100x magnifications respectively. c) AFM topography and d) height profile showing step-like height variation which corresponds to individual MoS$_2$ layers. e) Raman spectra for the multilayers at the nucleation center. f) PL mapping centered at 1.8eV shows the PL intensity drop at the center due to the thick layers and uniform outer region consists of monolayer MoS$_2$

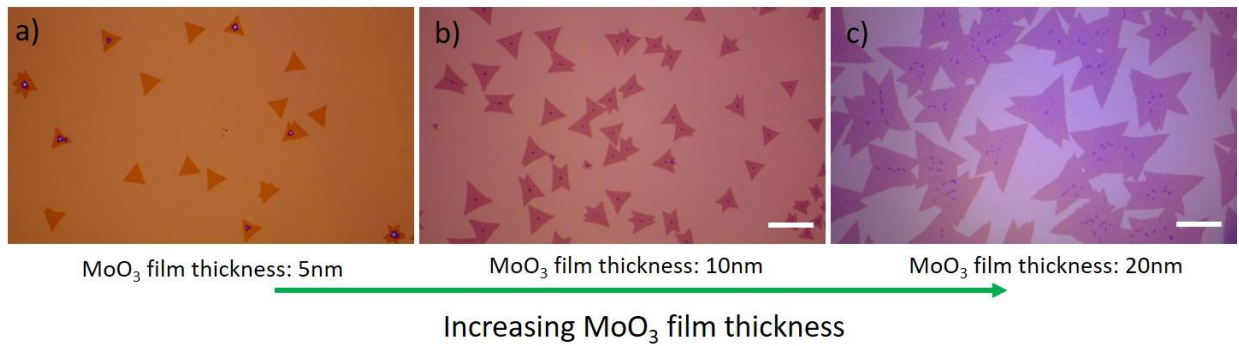

**Figure S5.** Effect of MoO3 film thickness. Optical micrographs of crystals grown with a)5nm, b) 10nm, c) 20nm MoO3 precursor thickness with 600mg of sulfur. Scale bar is 40µm in each image.

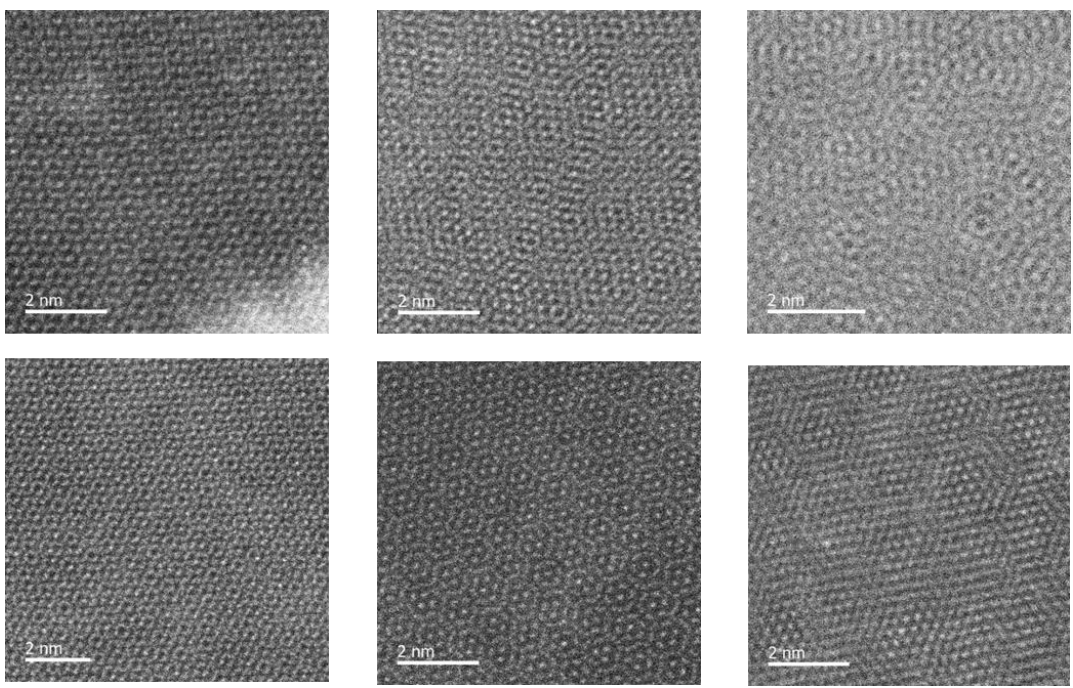

**Figure S6.** Supplementary TEM images. Different Moire patterns were seen at some locations of the transferred flakes depending on the angle of orientation of different layers at the polycrystalline regions and/or different folding orientations of the transferred films.

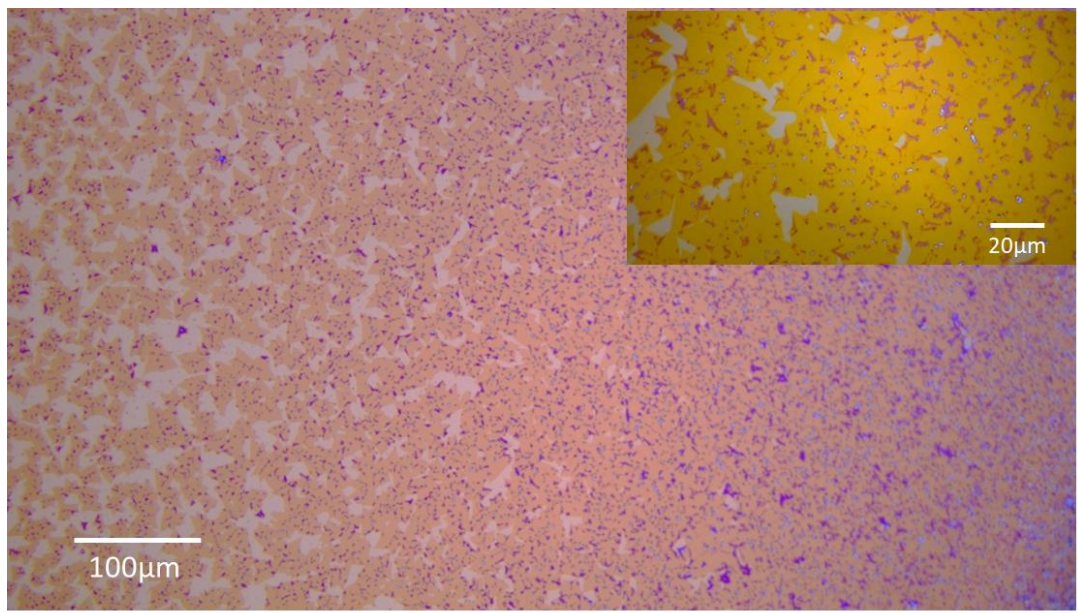

**Figure S7.** MoS$_2$ growth on source substrate. An optical micrograph of the 20nm MoO3 source substrate after the growth. An enlarged image is shown (100X) in the inset. We observed this growth at few regions of the source substrate and most of these regions are polycrystalline. Grain boundaries and overlapping is also visible in the optical images.

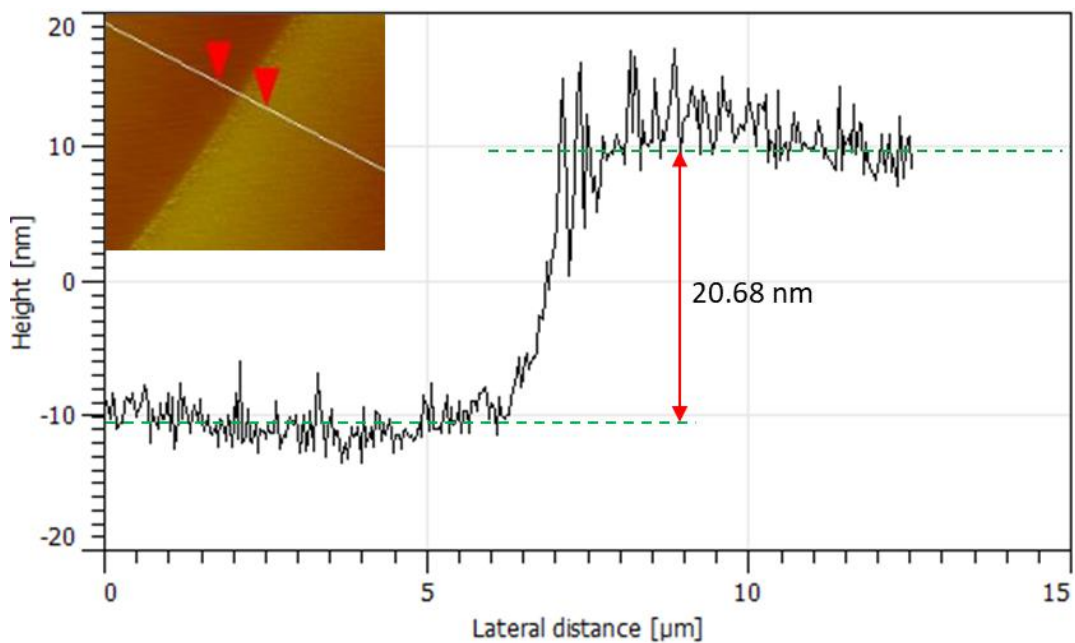

**Figure S8.** AFM characterization of thermally deposited thin film with target thickness of 20nm. AFM topography image shown in the inset express high uniformity of the deposited thin film. Height profile shows a clear step height of 20.68 nm at the substrate-thin film boundary